\documentclass[conference]{IEEEtran}
\IEEEoverridecommandlockouts
\usepackage{amsmath,amssymb,amsfonts}
\usepackage{graphicx}

\usepackage{algorithm}
\usepackage{algpseudocode}  

\usepackage{threeparttable}

\usepackage{subfigure}

\usepackage{multicol}

\usepackage{hyperref}
\usepackage{cleveref}

\begin{document}
	
	\title{A Novel Method to Estimate the Coordinates of LEDs in Wireless Optical Positioning Systems}
	
	\author{\IEEEauthorblockN{Kehan Zhang}
		\IEEEauthorblockA{School of Information Science\\ and Engineering\\
			Southeast University\\
			Nanjing, China\\
			Email: 220180736@seu.edu.cn}
		\and
		\IEEEauthorblockN{Zaichen Zhang}
		\IEEEauthorblockA{School of Information Science\\ and Engineering\\
			Southeast University\\
			Nanjing, China\\
			Email: zczhang@seu.edu.cn}
		\and
		\IEEEauthorblockN{Bingcheng Zhu}
		\IEEEauthorblockA{School of Information Science\\ and Engineering\\
			Southeast University\\
			Nanjing, China\\
			Email: zbc@seu.edu.cn}
		
		\thanks{Kehan Zhang, Zaichen Zhang and Bingcheng Zhu are with National Mobile Communications Research Laboratory, Southeast University, Nanjing 210096, China. Bingcheng Zhu is the corresponding author.}
		
		\thanks{This work is supported by NSFC projects (61960206005 and 61803211), the Fundamental Research Funds for the Central Universities (2242021k30043 and 2242021k30053), and Research Fund of National Mobile Communications Research Laboratory.}
		
	}
	
	\maketitle
	
\begin{abstract}
	
 Traditional visible light positioning (VLP) systems estimate receivers' coordinates based on the known light-emitting diode (LED) coordinates. However, the LED coordinates are not always known accurately. Because of the structural changes of the buildings due to temperature, humidity or material aging, even measured by highly accurate laser range finders, the LED coordinates may change unpredictably. In this paper, we propose an easy and low-cost method to update the position information of the LEDs. We use two optical angle-of-arrival (AOA) estimators to detect the beam directions of the LEDs. Each AOA estimator has four differently oriented photodiodes (PDs). Considering the additive noises of the PDs, we derive the closed-form error expression for the proposed LED coordinates estimator. Both analytical and Monte Carlo experimental results show that the layout of the AOA estimators could affect the estimation error. These results may provide intuitive insights for the design of the optical indoor positioning systems.
	
\end{abstract}

\IEEEpeerreviewmaketitle

\section{Introduction}	

VLP systems have aroused increasing attention from researchers for their high accuracy and low cost when light-of-sight channels exist. VLP algorithms had been roughly classified into three
categories \cite{Yuan}: received signal strength (RSS), time of arrival (TOA) or time difference of arrival (TDOA), and angle of arrival (AOA). 

Among different VLP algorithms, AOA has become one of the most popular topics. In the AOA-based systems, the transmitters or receivers are usually specially structured to detect the incident light directions, and these direction informations are combined to estimate the position of the receiver. In \cite{Yoshino}, an image sensor was exploited at the receiver to detect the signals from several beacon LEDs at the transmitter, and based on these signals, the receiver was localized. However, a drawback of image sensors is their narrow bandwidth, which hampers their applications in high rate visible light communication (VLC) systems. Reference \cite{Aparicio-Esteve, Cincotta} proposed a quadrant photodiode angular diversity aperture (QADA) as an AOA estimator to obtain the incident light direction to the receiver. In \cite{zhu}, the optimal arrangement of PDs was derived for the AOA estimator to achieve the incidence vector from the receiver to the beacon LED. Although realized through various techniques, aforementioned positioning systems demand accurate coordinate informations of the beacon LEDs.

In this paper, to cover the loophole of the traditional VLP systems, we aim to propose an efficient LED coordinates estimator. Such an estimator consists of two AOA estimators in \cite{zhu}, and the locations of AOA estimators are premeasured. Each AOA estimator could estimate the incident vector from the AOA estimator to the LED. The main contributions of this work can be summarized as follows:
\begin{itemize}
	\item An efficient LED localization system is proposed to avoid laborious measurement works with laser range finders.
	\item Both thermal noise and shot noise are considered to analyze the performance of the AOA estimator.
	\item Closed-form error expression is derived for the proposed LED localization system, providing a useful mathematical tool for the error analysis of such systems.
\end{itemize}

\section{System Model}

\subsection{Beacon LED Localization System}

\begin{figure}[htbp]
	\centerline{\includegraphics[scale=0.5]{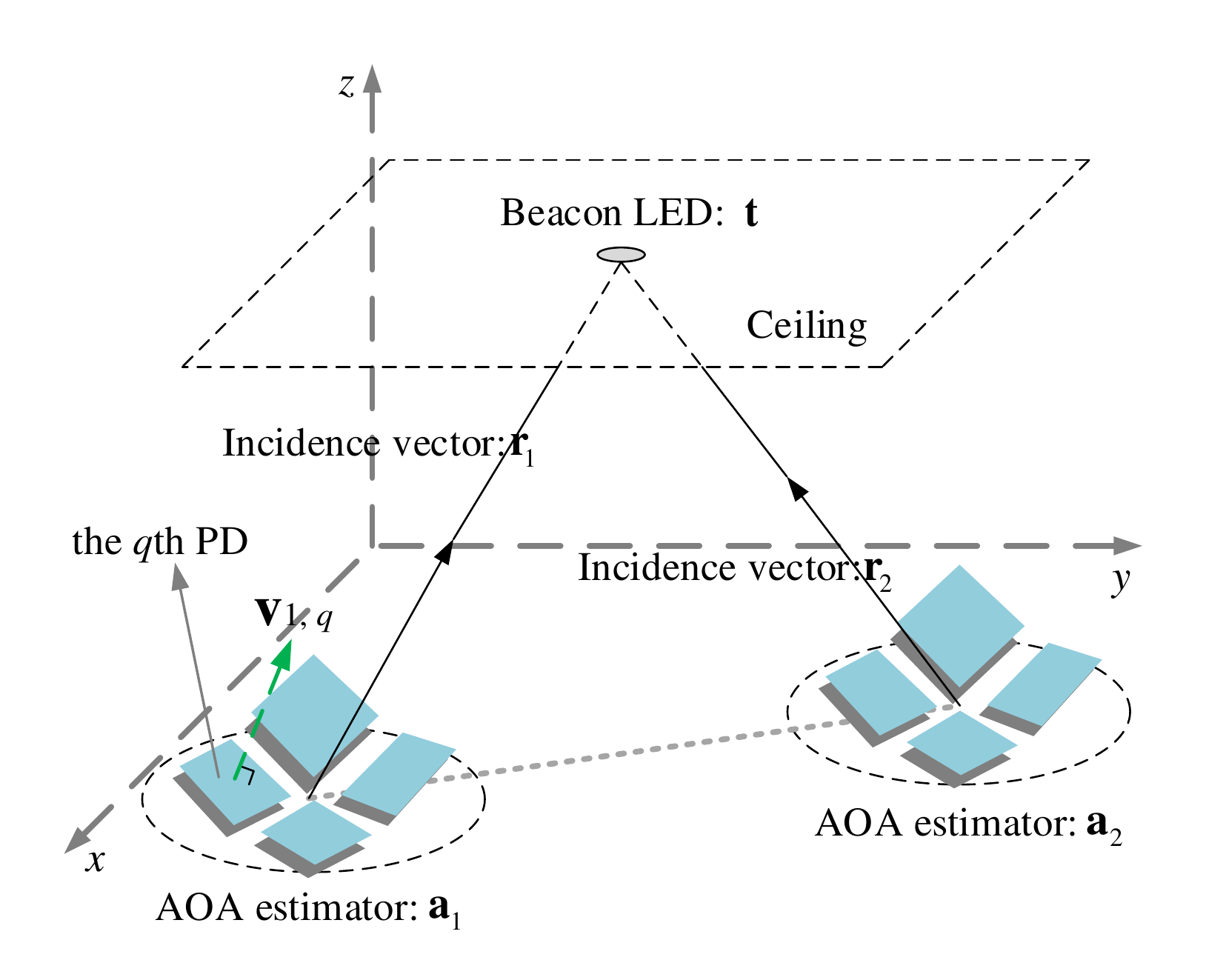}}
	\caption{Demonstration of the beacon LED localization system. The LED is placed on the ceiling whose unknown 3D coordinates are denoted by $\textbf{t}$. Two AOA estimators are set on the ground whose 3D coordinates are denoted by $\textbf{a}_1$ and $\textbf{a}_2$. The incidence vectors from the AOA estimators to the beacon LED are denoted by $\textbf{r}_1$ and $\textbf{r}_2$. Each AOA estimator consists of four PDs. The normalized normal vector of each PD is $\textbf{v}_{k,q}\ (k=1,2\ q=1,2,3,4)$.} 
	\label{model}
\end{figure}

As shown in Fig. \ref{model}, a beacon LED is placed on the ceiling with its three-dimensional (3D) coordinates denoted by a $3\times 1$ vector $\textbf{t}$. Two AOA estimators have their 3D coordinates at $\textbf{a}_1$ and $\textbf{a}_2$. Two normalized incidence vectors are oriented from the AOA estimators to the LED, which are $\textbf{r}_k=(\textbf{t}-\textbf{a}_k)/||\textbf{t}-\textbf{a}_k||\ (k=1,2)$. The AOA estimators could provide the incidence vector samples
$\hat{\textbf{r}}_1$ and $\hat{\textbf{r}}_2$, where $\hat{\textbf{r}}_k$ is the estimation of $\textbf{r}_k$. Therefore, the proposed localization system is to design a function $g(\cdot)$ that can output the estimation of the LED coordinates $\hat{\textbf{t}}$, i.e.
\begin{equation}
	\label{target}
	\hat{\textbf{t}} = g(\textbf{a}_1,\textbf{a}_2,\hat{\textbf{r}}_1,\hat{\textbf{r}}_2).
\end{equation}
The relation in (\ref{target}) will be developed in the latter context, and the estimation error $||\textbf{t} -\hat{\textbf{t}}||$ will also be analyzed.

\begin{figure}[htbp]
	\centerline{\includegraphics[scale=0.6]{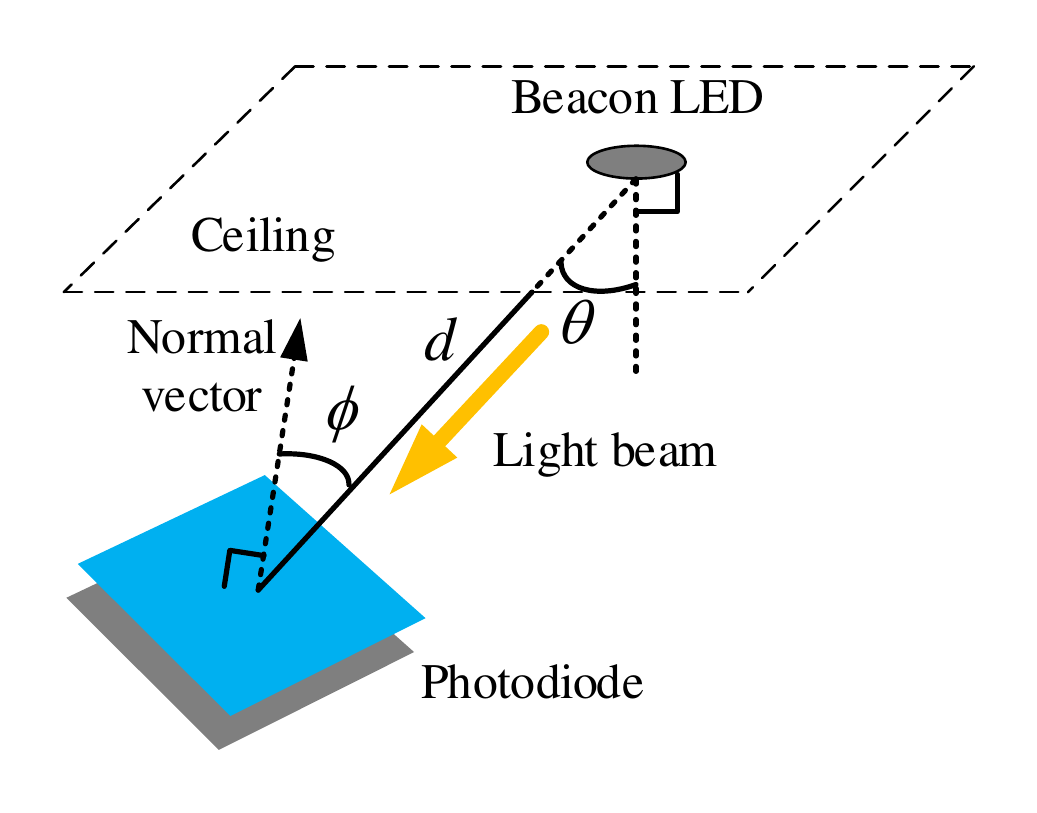}}
	\caption{Demonstration of Lambertian radiation model. $\theta$ is the radiation angle of the beacon LED and $\phi$ is the incidence angle of the PD. $d$ is the distance between the LED and the PD.}
	\label{lambert}
\end{figure}

\subsection{Lambertian Model and the Received Signal}

Lambertian model is a widely used model for describing the radiation of LEDs. As shown in Fig. \ref{lambert}, a beacon LED emits its light to a PD with the radiation angle $\theta$. The distance between the LED and the PD is $d$, and the incidence angle of the light beam is $\phi$. According to the Lambertian model, the received power is expressed as \cite{Pathak}
\begin{equation}
	\label{pr}
	P_r = P_t\frac{(m+1)s}{2\pi d^2}\cos^m\theta\cos\phi
\end{equation}
where $P_r$ is the received luminous flux of the PD; $P_t$ is the transmitted power of the LED; $s$ is the effective receiving area of the PD; $m$ is a constant parameter related to the LED. Considering the thermal noise and the shot noise \cite{komine}, the received signal power with noise $\hat{P}_r$ is
\begin{equation}
	\label{h_pt}
	\hat{P}_r = P_r + n
\end{equation}
where the noise $n$ is expressed in the form of luminance flux. The received optical power would be further converted into the electrical current by the PD, and the converted current signal with noise $\hat{\mu}$ is
\begin{equation}
	\label{h_mu}
	\begin{split}
		\hat{\mu} = \frac{R_p}{s}(P_r + n) = \mu + n^{\mu}
	\end{split}
\end{equation}
where $R_p$ is the conversion efficiency of the PD; $\mu = \frac{R_p}{s}P_r$; $n^{\mu} = n\frac{R_p}{s}$ is the noise of current. Referring to the minimum mean square error (MMSE)-fit noise model of \cite{yasir}, the variance of $n^{\mu}$ is expressed as
\begin{equation}
	\label{noise_}
	\begin{split}
		(\sigma^{\mu})^2
		=\ & 8.0185\times10^{-18} + 1.869\times10^{-11}\mu
	\end{split}
\end{equation}
where $(\sigma^{\mu})^2$ is the variance of $n^{\mu}$.

\subsection{AOA Estimator and the Error Vector of the Estimated Incidence Vector}

As shown in Fig. \ref{model}, the $k$th AOA estimator \cite{zhu} consists of four specially oriented PDs, and their normalized normal vectors are denoted by $3\times 1$ vectors $\textbf{v}_{k,q}\ (k=1,2\ \ q=1, 2, 3, 4)$. Such an AOA estimator is used to estimate the incidence vector $\textbf{r}_k$. Considering the noise in (\ref{h_pt}), the estimated incidence vector $\hat{\textbf{r}}_k$ is expressed as
\begin{equation}
	\label{hat_rk}
	\hat{\textbf{r}}_k = \textbf{r}_k + \textbf{n}_k,\ \ k=1,2
\end{equation}
where $\textbf{n}_k$ is the $3 \times 1$ error vector of the $k$th estimated incidence vector. To derive $\textbf{n}_k$, we need to review the process of the AOA estimation in \cite{zhu} and convert it to the current form. First  we could calculate the received light current $\hat{\mu}_{k,q}$ of the $q$th PD in the $k$th AOA estimator as
\begin{equation}
	\label{mu_kq}
	\begin{split}
		\hat{\mu}_{k,q}
		&=  R_pP_t\frac{m+1}{2\pi d_{k,q}^2}\cos^m\theta_{k,q} \textbf{v}_{k,q}^T\textbf{r}_{k,q}+ n^{\mu}_{k,q}
	\end{split}
\end{equation}
where $d_{k,q}$ is the distance from the beacon LED to the $q$th PD in the $k$th AOA estimator; $\theta_{k,q}$ is the radiation angle of the $q$th PD in the $k$th AOA estimator; $\textbf{r}_{k,q}$ is the normalized incident vector from the $q$th PD of the $k$th AOA estimator to the beacon LED; $n^{\mu}_{k,q}$ is the noise of the $q$th PD in the $k$th AOA estimator in the form of current.
Since the size of the AOA estimator is small, we could assume $d_{k,q}=d_k$, $\theta_{k,q}=\theta_k$ and $\textbf{r}_{k,q} = \textbf{r}_{k}$ \cite{Yang, zhu}, which means all the PDs in an AOA estimator share the same incidence vector. With this assumption, we could rewrite (\ref{mu_kq}) as
\begin{equation}
	\label{mukq}
	\begin{split}
		\hat{\mu}_{k,q} 
		&= \mu_{max,k}\textbf{v}_{k,q}^T\textbf{r}_{k} + n^{\mu}_{k,q}
	\end{split}
\end{equation}
where 
\begin{equation}
\mu_{max,k} = R_pP_t\frac{m+1}{2\pi d_k^2}\cos^m\theta_{k}.
\end{equation}
Combining the light current expressions for the four PDs in (\ref{mukq}), we obtain the received current vector
\begin{equation}
	\label{mu_matrix}
	\hat{\boldsymbol{\mu}}_k = \mu_{max,k}\textbf{V}_{PD}\textbf{r}_k + \textbf{n}_k^\mu
\end{equation}
where 
\begin{equation}
	\begin{split}
		\hat{\boldsymbol{\mu}}_k &= (\hat{\mu}_{k,1}, \hat{\mu}_{k,2}, \hat{\mu}_{k,3}, \hat{\mu}_{k,4})^T\\ \textbf{n}_k^\mu &= (n^{\mu}_{k,1}, n^{\mu}_{k,2}, n^{\mu}_{k,3}, n^{\mu}_{k,4})^T
	\end{split}
\end{equation}
and $\textbf{V}_{PD}$ is the normal vector matrix of the PDs as
\begin{equation}
	\textbf{V}_{PD} = (\textbf{v}_{k,1}, \textbf{v}_{k,2}, \textbf{v}_{k,3}, \textbf{v}_{k,4})^T.
\end{equation}
Applying the least square (LS) method to (\ref{mu_matrix}) \cite[eq. (5)]{zhu}, we could obtain
\begin{equation}
	\label{h_rk}
	\begin{split}
		\hat{\textbf{r}}_k
		&= \textbf{r}_k + \frac{1}{\mu_{max,k}}(\textbf{V}_{PD}^T\textbf{V}_{PD})^{-1}\textbf{V}_{PD}^T\textbf{n}_k^\mu.
	\end{split}
\end{equation}
Comparing (\ref{h_rk}) with (\ref{hat_rk}), $\textbf{n}_k$ is expressed as
\begin{equation}
	\label{n_k}
	\textbf{n}_k = \frac{1}{\mu_{max,k}}(\textbf{V}_{PD}^T\textbf{V}_{PD})^{-1}\textbf{V}_{PD}^T\textbf{n}_k^\mu.
\end{equation}
To minimize the average power of the AOA estimation error $E[||\textbf{n}_k||^2]$, $\textbf{V}_{PD}$ is optimized to \cite[eq. (20)]{zhu}
\begin{equation}
	\label{vpd_opt}
	\begin{split}
		\textbf{V}^*_{PD} =\ &
		\sqrt{\frac{2}{3}}
		\left (
		\begin{matrix}
			\cos\frac{\pi}{2}, &\sin\frac{\pi}{2}, &\frac{1}{\sqrt{2}}\\
			\cos\pi, &\sin\pi, &\frac{1}{\sqrt{2}}\\
			\cos\frac{3\pi}{2}, &\sin\frac{3\pi}{2}, &\frac{1}{\sqrt{2}}\\
			\cos2\pi, &\sin2\pi, &\frac{1}{\sqrt{2}}
		\end{matrix}
		\right )
	\end{split}
\end{equation}
where $\textbf{V}^*_{PD}$ is the optimal normal vector matrix when the shot noise is not considered. In this work, we adopt $\textbf{V}_{PD}^*$ as our normal vector matrix.

\section{LED Localization Algorithm}

The incidence vector $\textbf{r}_k$ is defined as\footnote{To clearly state the localization algorithm, noise is omitted in this section, but detailed discussion on the noise impact will be given in section V.}
\begin{equation}
	\label{rk}
	\begin{split}
		\textbf{r}_k = \frac{\textbf{t}-\textbf{a}_k}{||\textbf{t}-\textbf{a}_k||} 
		= \frac{\textbf{t}-\textbf{a}_k}{d_k}\ \ k = 1,2
	\end{split}
\end{equation}
where $d_k = ||\textbf{t}-\textbf{a}_k||$ is the distance from the $k$th AOA estimator to the beacon LED, while $\textbf{r}_k$, $\textbf{a}_k$ and $\textbf{t}$ are all $3\times 1$ vectors expressed as
\begin{equation}
	\label{vs}
	\textbf{r}_k =
	\left (
	\begin{matrix}
		r_{k,1} \\
		r_{k,2} \\
		r_{k,3}
	\end{matrix}
	\right  ),\
	\textbf{a}_k =
	\left (
	\begin{matrix}
		a_{k,1} \\
		a_{k,2} \\
		a_{k,3}
	\end{matrix}
	\right  ),\
	\textbf{t} =
	\left (
	\begin{matrix}
		t_{1} \\
		t_{2} \\
		t_{3}
	\end{matrix}
	\right  ).
\end{equation}
From (\ref{rk}), we obtain
\begin{equation}
	\label{u}
	\textbf{t} = \textbf{a}_1 + d_1\textbf{r}_1 = \textbf{a}_2 + d_2\textbf{r}_2
\end{equation}
which can be transformed to
\begin{equation}
	\label{matrix_I}
	\left (
	\begin{matrix}
		\textbf{r}_1, \textbf{r}_2
	\end{matrix}
	\right  )
	\left (
	\begin{matrix}
		d_1 \\
		-d_2
	\end{matrix}
	\right ) = \textbf{a}_2 - \textbf{a}_1.
\end{equation}
With the LS method, we get
\begin{equation}
	\label{am_an}
	\left (
	\begin{matrix}
		d_1 \\
		-d_2
	\end{matrix}
	\right ) 
	= (\textbf{A}^T\textbf{A})^{-1}\textbf{A}^T(\textbf{a}_2 - \textbf{a}_1)
\end{equation}
where
\begin{equation}
	\begin{split}
		\textbf{A} &= 
		\left (
		\begin{matrix}
			\textbf{r}_1, \textbf{r}_2
		\end{matrix}
		\right  )
	\end{split}
\end{equation}
and $\textbf{A}^T\textbf{A}$ is calculated as
\begin{equation}
	\begin{split}
		\textbf{A}^T\textbf{A} = 
		\left (
		\begin{matrix}
			\textbf{r}_1^T\textbf{r}_1,\ \ \textbf{r}_1^T\textbf{r}_2\\
			\textbf{r}_2^T\textbf{r}_1,\ \ \textbf{r}_2^T\textbf{r}_2
		\end{matrix}
		\right ) 
		=
		\left (
		\begin{matrix}
			c_1,\ c_2 \\
			c_2,\ c_3
		\end{matrix}
		\right )
	\end{split}
\end{equation}
where $c_1$, $c_2$ and $c_3$ are
\begin{equation}
	\label{c}
	c_1 = \textbf{r}_1^T\textbf{r}_1,\
	c_2 =\textbf{r}_1^T\textbf{r}_2=\textbf{r}_2^T\textbf{r}_1,\
	c_3 = \textbf{r}_2^T\textbf{r}_2
\end{equation}
and the inverse matrix of $\textbf{A}^T\textbf{A}$ is obtained as
\begin{equation}
	\label{ATA}
	(\textbf{A}^T\textbf{A})^{-1} = \frac{1}{c_1c_3-c_2^2}
	\left (
	\begin{matrix}
		c_3,\ -c_2\\
		-c_2,\ c_1
	\end{matrix}
	\right).
\end{equation}
The rest part of (\ref{am_an}), i.e. $\textbf{A}^T(\textbf{a}_2-\textbf{a}_1)$, is calculated as
\begin{equation}
	\label{ATa}
	\begin{split}
		\textbf{A}^T(\textbf{a}_2-\textbf{a}_1) =
		\left (
		\begin{matrix}
			\textbf{r}_1^T(\textbf{a}_2-\textbf{a}_1)\\
			\textbf{r}_2^T(\textbf{a}_2-\textbf{a}_1)
		\end{matrix}
		\right )
		=
		\left (
		\begin{matrix}
			f_1\\
			f_2
		\end{matrix}
		\right )
	\end{split}
\end{equation}
where $f_1$ and $f_2$ are
\begin{equation}
	\label{f}
	\begin{aligned}
		f_k = \textbf{r}_k^T(\textbf{a}_2-\textbf{a}_1),\ \ k=1,2.
	\end{aligned}
\end{equation} 
Together with (\ref{am_an}), (\ref{ATA}) and (\ref{ATa}), $d_1$ and $d_2$ are expressed as
\begin{equation}
	\label{result}
	\left\{
	\begin{aligned}
		d_1 &= \frac{c_3f_1-c_2f_2}{c_1c_3-c_2^2}\\
		d_2 &= \frac{c_2f_1-c_1f_2}{c_1c_3-c_2^2}.
	\end{aligned}
	\right.
\end{equation}
We could derive the estimation of $\textbf{t}$ with (\ref{u}) and (\ref{am_an}) as
\begin{equation}
	\label{t}
	\begin{split}
		\hat{\textbf{t}} =&\ \frac{\textbf{a}_1+d_1\textbf{r}_1+\textbf{a}_2+d_2\textbf{r}_2}{2}\\
		=&\ \frac{\textbf{a}_1+\textbf{a}_2}{2} + \frac{\textbf{A}}{2} 
		\left (
		\begin{matrix}
			1, &0\\
			0, &-1
		\end{matrix}
		\right )
		(\textbf{A}^T\textbf{A})^{-1}\textbf{A}^T(\textbf{a}_2 - \textbf{a}_1)
	\end{split}
\end{equation}
where $\hat{\textbf{t}}$ is the estimation of $\textbf{t}$, $\textbf{A}$ is the output of the AOA estimators, and $\textbf{a}_1$, $\textbf{a}_2$ are the known AOA estimators' positions.

\section{Error Analysis}

According to (\ref{n_k}) and (\ref{vpd_opt}), $\textbf{n}_k$ is calculated as
\begin{equation}
	\label{tnk}
	\begin{split}
		\textbf{n}_{k} =
		\left (
		\begin{matrix}
			n_{k,1}\\
			n_{k,2}\\
			n_{k,3}
		\end{matrix}
		\right )
		 = \frac{\sqrt{6}}{4\mu_{max,k}}
		\left (
		\begin{matrix}
			n^\mu_{k,4}-n^\mu_{k,2}\\
			n^\mu_{k,1}-n^\mu_{k,3}\\
			\frac{1}{\sqrt{2}}\sum_{q=1}^4 n^\mu_{k,q}
		\end{matrix}
		\right ).
	\end{split}
\end{equation}
where $n^\mu_{k,1}$, $n^\mu_{k,2}$, $n^\mu_{k,3}$ and $n^\mu_{k 4}$ are independent. We could calculate the auto-covariance matrix $\textbf{C}_{\textbf{n}_k} = 	E\big\{\textbf{n}_k\textbf{n}_k^T\big\}$ as
\begin{equation}
	\label{auto_co}
	\textbf{C}_{\textbf{n}_k} = 
	\left (
	\begin{matrix}
		&E[n^2_{k,1}], &0, &E[n_{k,1}n_{k,3}]\ \ \ \\
		&0, &E[n^2_{k,2}], &E[n_{k,2}n_{k,3}]\ \ \ \\
		&E[n_{k,3}n_{k,1}], &E[n_{k,3}n_{k,2}], &E[n^2_{k,3}]\ \ \
	\end{matrix}
	\right ).
\end{equation}
Since $\textbf{a}_1$ and $\textbf{a}_2$ are fixed parameters, we should rewrite (\ref{target}) as
\begin{equation}
	\label{target_new}
	\hat{\textbf{t}} = g(\hat{\textbf{r}}_1,\hat{\textbf{r}}_2).
\end{equation}
According to (\ref{noise_}) and (\ref{tnk}), the noise $\textbf{n}_k$ could be assumed as a tiny disturbance to the incidence vector $\textbf{r}_k$. Therefore, we could apply Taylor series expansion to (\ref{target_new}) at point $\textbf{t}$, where $\hat{\textbf{r}}_k = \textbf{r}_k$,
\begin{equation}
	\label{tay}
	\begin{split}
		\hat{\textbf{t}} &= g(\textbf{r}_1, \textbf{r}_2) + \sum_{k=1}^2 \frac{\partial g(\hat{\textbf{r}}_1,\hat{\textbf{r}}_2)}{\partial \hat{\textbf{r}}_k}\bigg|_{\hat{\textbf{r}}_k = \textbf{r}_k} (\hat{\textbf{r}}_k - \textbf{r}_k)+\cdots\\
		&\approx \textbf{t} + \sum_{k=1}^2 \frac{\partial \hat{\textbf{t}}}{\partial \textbf{r}_k} \textbf{n}_k.
	\end{split}
\end{equation}
Subtracting $\textbf{t}$ from both sides of (\ref{tay}), the estimation error $\textbf{e}_r$ is obtained as
\begin{equation}
	\label{dt}
	\begin{split}
		\textbf{e}_r = 
		\hat{\textbf{t}} - \textbf{t}
		\approx \sum_{k=1}^2 \frac{\partial \hat{\textbf{t}}}{\partial \textbf{r}_k} \textbf{n}_k
	\end{split}
\end{equation}
where
\begin{equation}
	\begin{split}
			\textbf{e}_r &= (e_{r,1}, e_{r,2}, e_{r,3})^T,\\
	\end{split}
\end{equation}
Equation (\ref{dt}) reveals the relation between the estimation error $\textbf{e}_r$ and the noise $\textbf{n}_k$. The derivatives in (\ref{dt}) essentially determines how robust the LED coordinate estimator is to the additive noise. Equation (\ref{t}) leads to
\begin{equation}
	\label{par}
	\begin{split}
		\frac{\partial \hat{\textbf{t}}}{\partial \textbf{r}_k}
		&= \frac{1}{2} \bigg(\textbf{r}_1\frac{\partial d_1}{\partial \textbf{r}_k} + d_1\frac{\partial \textbf{r}_1}{\partial \textbf{r}_k} +
		\textbf{r}_2\frac{\partial d_2}{\partial \textbf{r}_k} + d_2\frac{\partial \textbf{r}_2}{\partial \textbf{r}_k}\bigg) \\
		&= \frac{1}{2} \bigg(\textbf{r}_1\frac{\partial d_1}{\partial \textbf{r}_k} + d_k\textbf{I}_3 +
		\textbf{r}_2\frac{\partial d_2}{\partial \textbf{r}_k}\bigg)
	\end{split}
\end{equation}
where $\textbf{I}_3$ is the $3\times 3$ unit matrix.
From (\ref{result}), we could calculate $\frac{\partial d_1}{\partial \textbf{r}_1}$ and $\frac{\partial d_2}{\partial \textbf{r}_1}$ as
\begin{equation}
	\label{par_1}
	\begin{split}
		\frac{\partial d_1}{\partial \textbf{r}_1} 
		=\ & -\frac{c_3f_1-c_2f_2}{(c_1c_3-c_2^2)^2}(2c_3\textbf{r}_1^T-2c_2\textbf{r}_2^T) \\
		&+ \frac{1}{c_1c_3-c_2^2}\bigg[c_3(\textbf{a}_2-\textbf{a}_1)^T-f_2\textbf{r}_2^T\bigg]\\
		\frac{\partial d_2}{\partial \textbf{r}_1} 
		=\ & -\frac{c_2f_1-c_1f_2}{(c_1c_3-c_2^2)^2}(2c_3\textbf{r}_1^T-2c_2\textbf{r}_2^T)\\
		&+\frac{1}{c_1c_3-c_2^2}\bigg[c_2(\textbf{a}_2-\textbf{a}_1)^T+f_1\textbf{r}_2^T-2f_2\textbf{r}_1^T\bigg].
	\end{split}
\end{equation}
Similarly, $\frac{\partial d_1}{\partial \textbf{r}_2}$ and $\frac{\partial d_2}{\partial \textbf{r}_2}$ are calculated as
\begin{equation}
	\label{par_2}
	\begin{split}
		\frac{\partial d_1}{\partial \textbf{r}_2} 
		=\ & -\frac{c_3f_1-c_2f_2}{(c_1c_3-c_2^2)^2} \bigg(2c_1\textbf{r}_2^T-2c_2\textbf{r}_1^T\bigg) \\
		&+ \frac{1}{c_1c_3-c_2^2} \bigg[ 2f_1\textbf{r}_2^T - c_2(\textbf{a}_2-\textbf{a}_1)^T-f_2\textbf{r}_1^T \bigg]\\
		\frac{\partial d_2}{\partial \textbf{r}_2} 
		=\ & -\frac{c_2f_1-c_1f_2}{(c_1c_3-c_2^2)^2} \bigg( 2c_1\textbf{r}_2^T  - 2c_2 \textbf{r}_1^T \bigg) \\
		&+
		\frac{1}{c_1c_3-c_2^2} \bigg[ f_1\textbf{r}_1^T - c_1(\textbf{a}_2-\textbf{a}_1)^T \bigg].
	\end{split}
\end{equation}
By substituting (\ref{par_1}) and (\ref{par_2}) into (\ref{par}), we could derive $\frac{\partial \hat{\textbf{t}}}{\partial \textbf{r}_1}$ and $\frac{\partial \hat{\textbf{t}}}{\partial \textbf{r}_2}$. With (\ref{tnk}), (\ref{par}), (\ref{par_1}) and (\ref{par_2}), we could calculate (\ref{dt}) and achieve the auto-covariance matrix of $\textbf{e}_r$ as

\begin{equation}
	\label{var1}
	\begin{split}
		E\{\textbf{e}_r\textbf{e}_r^T\} 
		\approx\ &
		E\bigg\{\big(\frac{\partial \hat{\textbf{t}}}{\partial \textbf{r}_1}\textbf{n}_1 + \frac{\partial \hat{\textbf{t}}}{\partial \textbf{r}_2}\textbf{n}_2\big)\big(\frac{\partial \hat{\textbf{t}}}{\partial \textbf{r}_1}\textbf{n}_1 + \frac{\partial \hat{\textbf{t}}}{\partial \textbf{r}_2}\textbf{n}_2\big)^T\bigg\} \\
		=\ & \frac{\partial \hat{\textbf{t}}}{\partial \textbf{r}_1}
		\textbf{C}_{\textbf{n}_1}\big(\frac{\partial \hat{\textbf{t}}}{\partial \textbf{r}_1}\big)^T +
		\frac{\partial \hat{\textbf{t}}}{\partial \textbf{r}_2}
		\textbf{C}_{\textbf{n}_2}\big(\frac{\partial \hat{\textbf{t}}}{\partial \textbf{r}_2}\big)^T.
	\end{split}
\end{equation}
Therefore, the theoretical LED positioning error $e_{ps}$ could be denoted by
\begin{equation}
	\label{error}
	\begin{split}
		e_{ps} &= \sqrt{E[e_{r,1}^2] + E[e_{r,2}^2] + E[e_{r,3}^2]}
		= \sqrt{tr(E\{\textbf{e}_r\textbf{e}_r^T\})}\\
		&\approx \sqrt{tr\bigg(\frac{\partial \hat{\textbf{t}}}{\partial \textbf{r}_1}
			\textbf{C}_{\textbf{n}_1}\big(\frac{\partial \hat{\textbf{t}}}{\partial \textbf{r}_1}\big)^T +
			\frac{\partial \hat{\textbf{t}}}{\partial \textbf{r}_2}
			\textbf{C}_{\textbf{n}_2}\big(\frac{\partial \hat{\textbf{t}}}{\partial \textbf{r}_2}\big)^T}\bigg)
	\end{split}
\end{equation}
where $tr(\cdot)$ denotes the trace of a matrix. 
Equation (\ref{error}) is a novel and important equation that relate the noise statistics to the LED localization error at a certain point. With this new analytical tool, we could avoid resorting to time-consuming Monte Carlo simulations and evaluate the system error effectively.

\section{Simulation Result}

We use the estimated incidence vector $\hat{\textbf{r}}_k$ in (\ref{hat_rk}) as the input of the LED localization algorithm in (\ref{t}).
To test the proposed system, a room of $4\times 4\times 4\ {\rm m^3}$ is constructed, with a beacon LED on the ceiling and two AOA estimators on the ground. In other words, the $z$-coordinates of the beacon LED and the AOA estimators are 4 m and 0 m, respectively. Each AOA estimator has four differently oriented PDs. The transmit power $P_t=5000\ {\rm lm}$, the effective receiving area of each PD $s = 15\ {\rm mm}^2$, the conversion efficiency $R_p = 22 {\rm nA/lux}$, and the constant $m=1$.

\begin{figure}[htbp]
	\centering
	\subfigure[Theoretical calculations of $e_{ps}$.]{
		\includegraphics[scale=0.5]{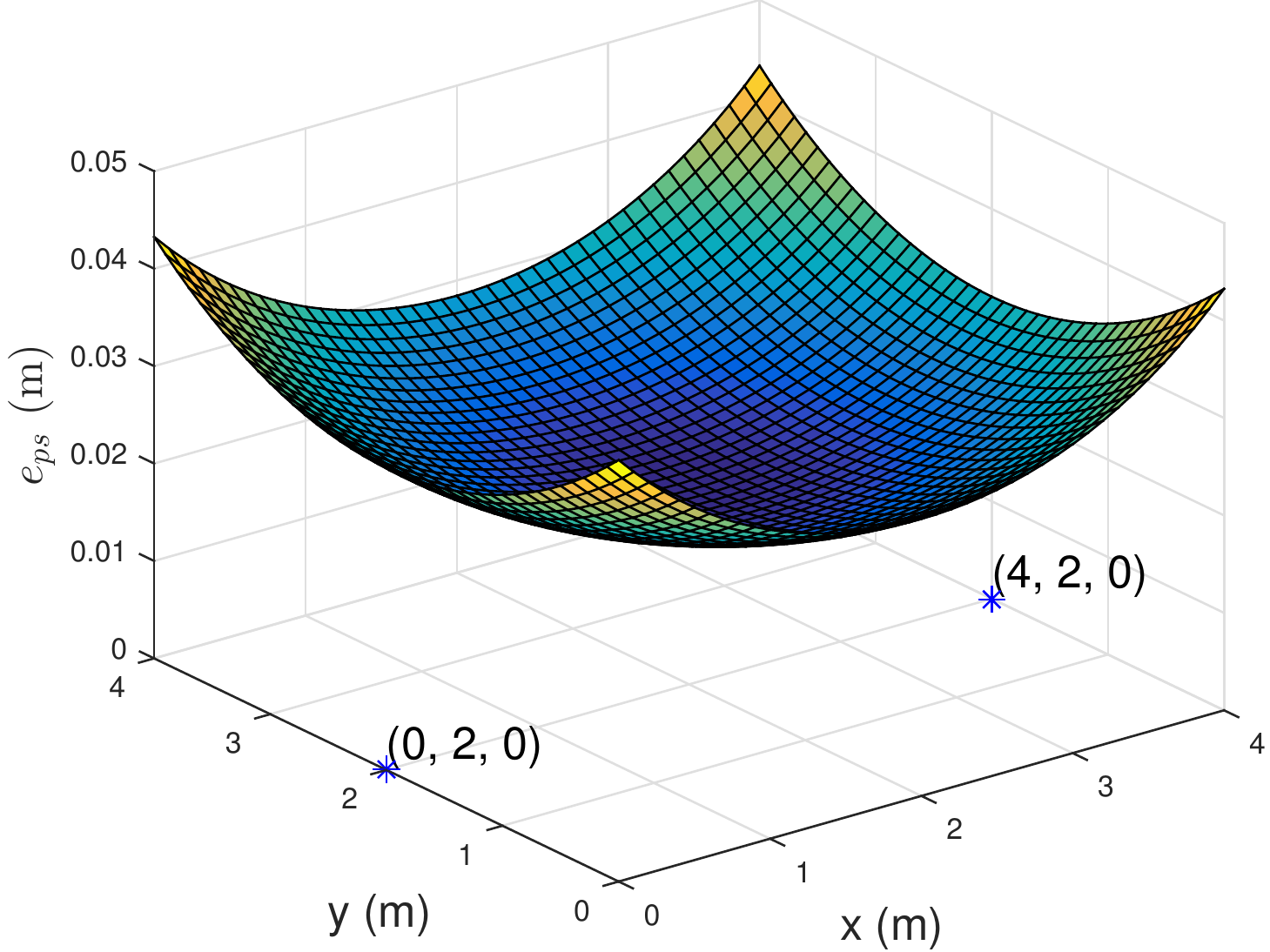}
		\label{3_a}
	}
	\quad
	\subfigure[Simulation results of $e_{ps}$.]{
		\includegraphics[scale=0.5]{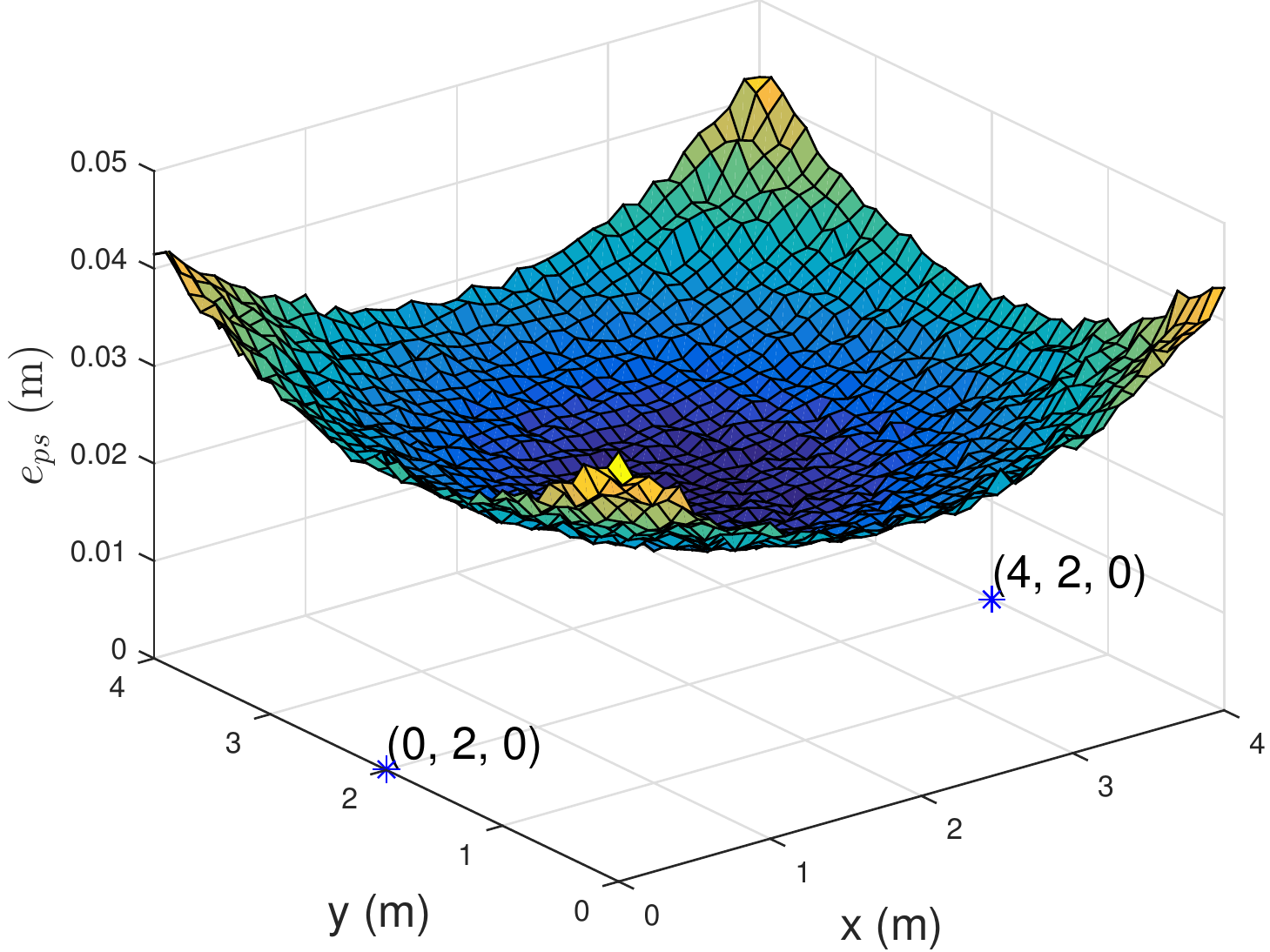}
		\label{3_b}
	}
	\caption{Theoretical calculations and simulation results of $e_{ps}$ are plotted in subfigure (a) and subfigure (b) respectively. Two AOA estimators are set at $(0\ {\rm m}, 2\ {\rm m}, 0\ {\rm m})^T$ and $(4\ {\rm m}, 2\ {\rm m}, 0\ {\rm m})^T$. The blue stars denote the AOA estimators, whose 3D coordinates are marked in the figure. $x$ and $y$ coordinates represent the horizontal coordinates of the beacon LED.}
	\label{3}
\end{figure}

\begin{figure}[htbp]
	\centering
	\subfigure[Theoretical calculations of $e_{ps}$.]{
		\includegraphics[scale=0.5]{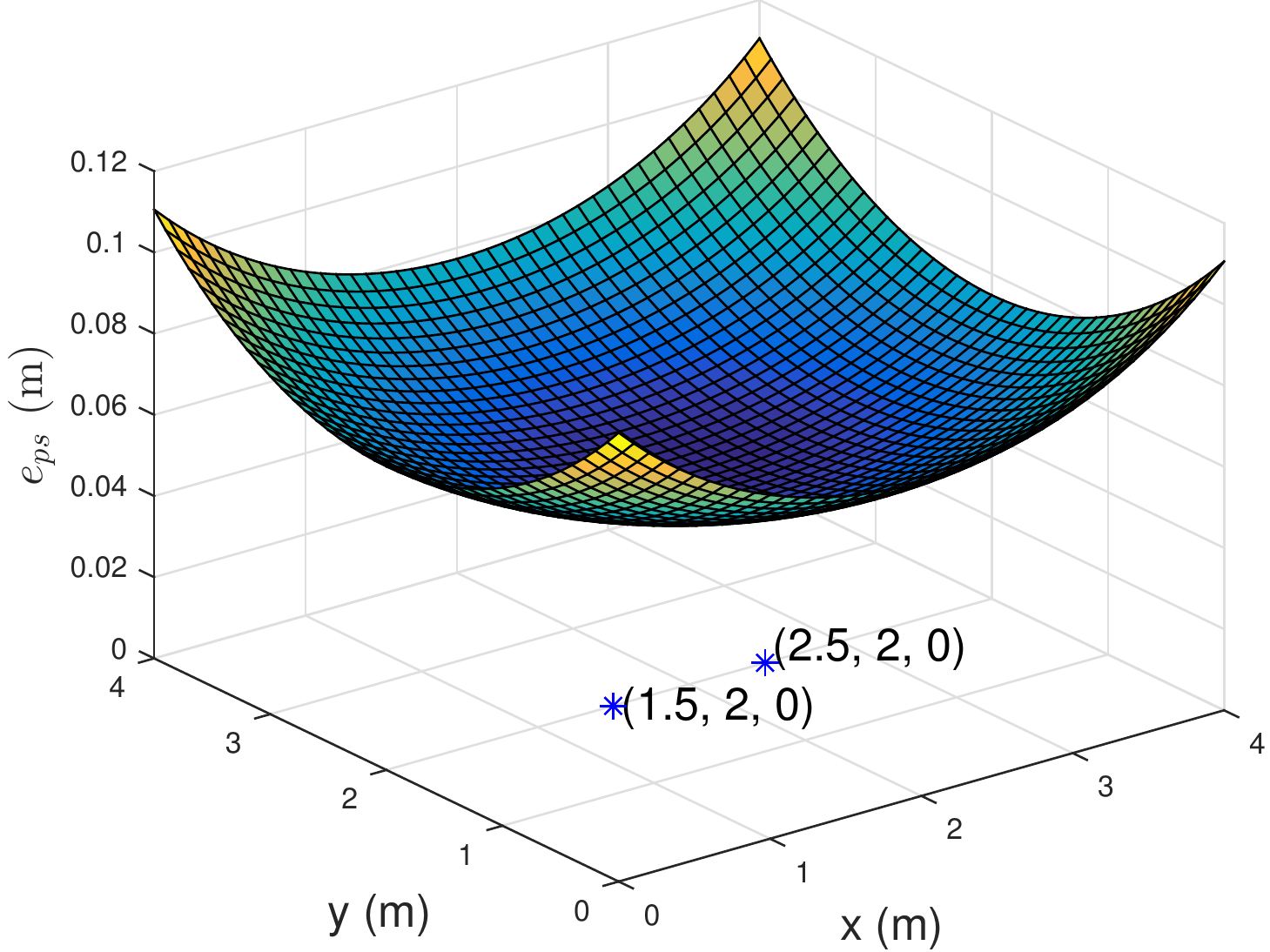}
		\label{1_a}
	}
	\quad
	\subfigure[Simulation results of $e_{ps}$.]{
		\includegraphics[scale=0.5]{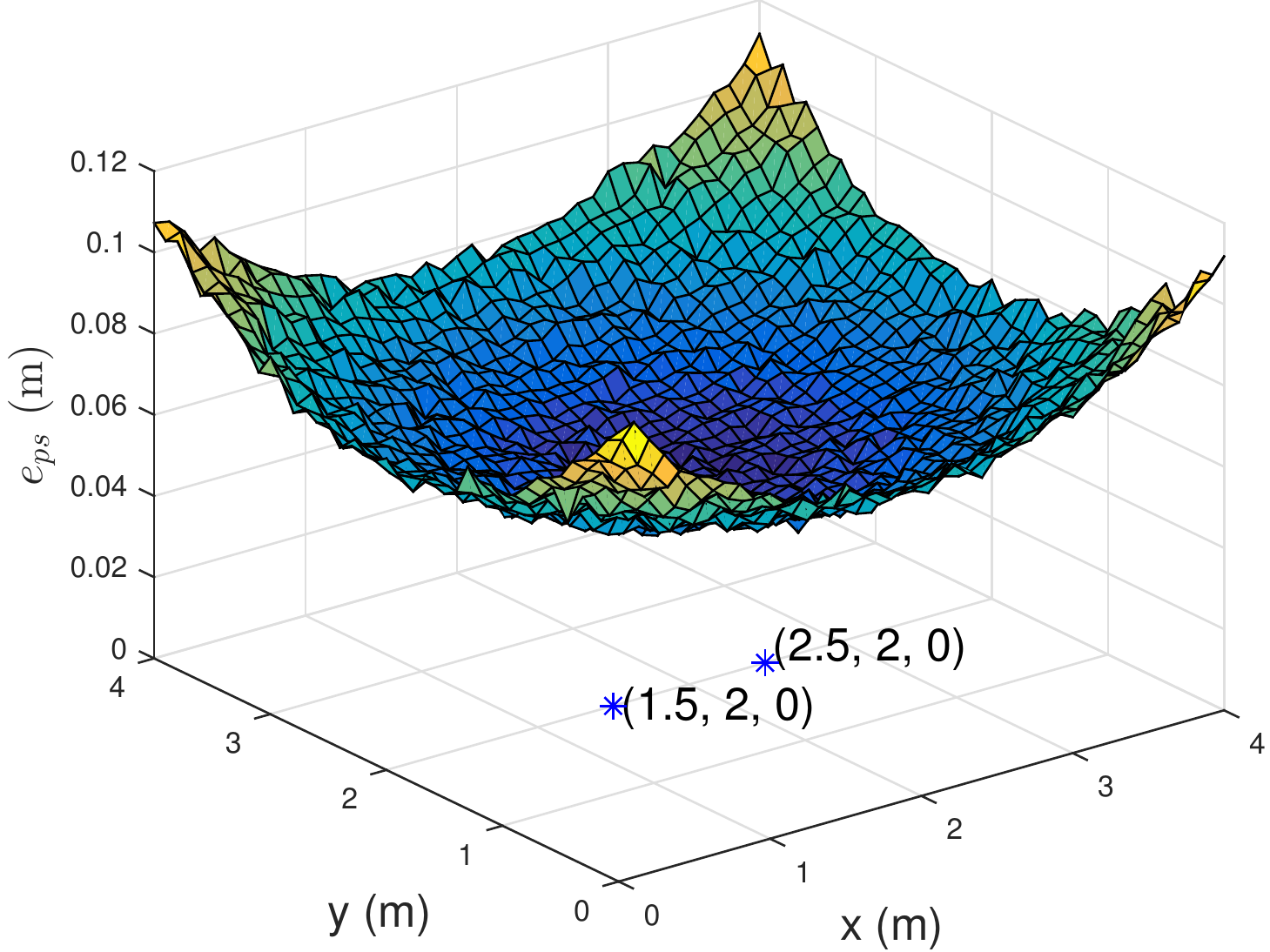}
		\label{1_b}
	}
	\caption{Theoretical calculations and simulation results of $e_{ps}$ are plotted in subfigure (a) and subfigure (b) respectively. Two AOA estimators are set at $(1.5\ {\rm m}, 2\ {\rm m}, 0\ {\rm m})^T$ and $(2.5\ {\rm m}, 2\ {\rm m}, 0\ {\rm m})^T$. The blue stars denote the AOA estimators, whose 3D coordinates are marked in the figure. $x$ and $y$ coordinates represent the horizontal coordinates of the beacon LED.}
	\label{1}
\end{figure}

In this section, we would investigate whether the result in (\ref{error}) agrees with the simulation results from the Monte Carlo experiments. We use the noise model in (\ref{noise_}), and prepare two typical placements of the AOA estimators. Then, we plot the theoretical calculations of $e_{ps}$ in subfigures (a) of Figs. \ref{3}-\ref{1}, and plot the Monte Carlo simulation results of $e_{ps}$ in subfigures (b).

As shown in Figs. \ref{3}-\ref{1}, the theoretical calculations and the Monte Carlo simulation results fit well, which proves the derivation in (\ref{noise_}) to be correct. In these two figures, $e_{ps}$ is minimized between the AOA estimators, which counters the intuition that the error is minimized right above the AOA estimators where the received power of one AOA estimator reaches its maximum. As shown in Fig. \ref{3}, the maximum positioning error $e_{ps}$ is below 5 cm, indicating low overall estimation error under such placement of the AOA estimators. In Fig. \ref{1}, two AOA estimators are placed closer, and the maximum estimation error is over 10 cm. Under such arrangement of the AOA estimators, poor overall localization performance turns out. A feasible explanation for this is that, when the AOA estimators are close, $\textbf{a}_1\approx\textbf{a}_2$, thus $c_1c_3\approx c_2^2$ as defined in (\ref{c}). Therefore, the factor $1/(c_1c_3-c_2^2)$ in (\ref{par_1}) and (\ref{par_2}) increases greatly, which brings an unwanted magnification for $e_{ps}$.

In summary, we should be cautious about the placement of the AOA estimators. Referring to the $e_{ps}$ in (\ref{error}), the two AOA estimators should not be set too close to each other.

\section{Conclusion}

In this paper, a beacon LED localization system is proposed to reduce the uncertainty of the LED positions. Two AOA estimators are exploited to estimate the incidence vectors from the AOA estimators to the LED. With the incidence vectors and the positions of the AOA estimators, the proposed system could localize the LEDs. To evaluate the estimation error of the system, we derive a closed-form error expression in terms of both thermal noise and shot noise.
Based on the analytical error results, the placement of the AOA estimators could affect the error greatly, and a reasonable advice for the placement is that the two AOA estimators should not be set too close to each other. With this proposed system, we have a new approach to build beacon LED coordinates databases without laborious measurements, and the mathematical tool quantifies the localization error and provides insights into the indoor optical positioning system design.

\end{document}